\documentstyle[wstwocl,epsfig]{article}
\begin{document}

\bibliographystyle{unsrt}    

\hyphenation{saddle tri-vi-al-ly}
\newcommand{\gtapprox}
{{\,}^{\raisebox{-.6ex}{${\scriptstyle>}$}}_
{\raisebox{.6ex}{${\scriptstyle\sim}$}}\,}
\newcommand{\ltapprox}
{{\,}^{\raisebox{-.6ex}{${\scriptstyle<}$}}_
{\raisebox{.6ex}{${\scriptstyle\sim}$}}\,}
\newcommand{\Bslash}{{\not{}B}}
\newcommand{\Dslash}{{D\hspace{-.7em}/}}
\newcommand{\st}{\scriptstyle}
\newcommand{\sst}{\scriptscriptstyle}
\newcommand{\mco}{\multicolumn}
\newcommand{\epp}{\epsilon^{\prime}}
\newcommand{\vep}{\varepsilon}
\newcommand{\ra}{\rightarrow}
\newcommand{\ppg}{\pi^+\pi^-\gamma}
\newcommand{\vp}{{\bf p}}
\newcommand{\ko}{K^0}
\newcommand{\kb}{\bar{K^0}}
\newcommand{\al}{\alpha}
\newcommand{\ab}{\bar{\alpha}}
\def\be{\begin{equation}}
\def\ee{\end{equation}}
\def\bea{\begin{eqnarray}}
\def\eea{\end{eqnarray}}
\def\CPbar{\hbox{{\rm CP}\hskip-1.80em{/}}}

\title{A ${\bf Z}_2$ Origin for B+L Violation in the Hot Electroweak Theory}

\firstauthors{Holger Bech Nielsen}

\firstaddress{The Niels Bohr Institute,\\ Blegdamsvej 17, DK-2100 Copenhagen,
Denmark}

\secondauthors{Minos Axenides$^a$, Andrei Johansen$^b$,
Ola T\"{o}rnkvist$^{c\,\ast}$}

\secondaddress{$^a$Dept.\ of Physics,
University of Crete, GR-71409 Iraklion, Greece\\
$^b$Lyman Laboratory of Physics,\\ Harvard University, Cambridge, MA
02138, USA\\
$^c$Nordita, Blegdamsvej 17, DK-2100 Copenhagen, Denmark}




\twocolumn[\maketitle\abstracts{ The space of static finite-energy
configurations of the electroweak theory admits a ${\bf Z}_{2}$
topological structure.
Odd-parity configurations with odd pure-gauge behavior at spatial infinity
(S sphaleron, W-Z strings, multisphalerons) mediate rapid $B+L$
violation
in the Early Universe.
Configurations with even pure-gauge behavior (S$^{*}$ sphalerons,
W-Z strings, multisphalerons)
are trivial and topologically equivalent to the vacuum.}]

The existence of odd-parity saddle point solutions (sphalerons) to the field
equations of the bosonic sector of the Weinberg-Salam theory has raised the
possibility that the baryon asymmetry of the Universe $( n_b /s
\sim 10^{-10})$
was generated in the electroweak phase transition $( T_{c}
\sim 100-300\ {\rm GeV})$\,\cite {Ku}.
To that effect, one must require that the $B$-violation rate induced
by the sphaleron in the broken phase $(M_w \ltapprox T < T_{c})$
increases with temperature
$(\Gamma_{B}
\sim \exp \{-M_w(T)/T\})$, as well
as the presence of unsuppressed $B$ violation in the symmetric phase
$(\Gamma_{B}
\sim \alpha_w T^4$, $T \geq T_{c})$. It is believed that for $M_w \ltapprox
T \ltapprox T_{c}$
the perturbation expansion around the sphaleron is not valid
because it yields a prefactor to the $B$-violation rate which
vanishes at $T=T_{c}$ \cite{Dia}.
 Sphaleron-like
configurations (deformations) are expected to dominate the $B$-violation
rate from
there on and into the symmetric phase. In the context of the standard $SU(2)
\times
U(1)$ theory we have characterized such configurations in terms of their
gauge-invariant generalized odd parity property
$$
A_i(-x)=-A^{S}_{i}(x)\equiv
$$
\be
-[S(x)A_{i}(x)S^{-1}(x) - i\partial_{i}S(x)S^{-1}(x)]
\ee
for some $S \in SU(2) \times U(1)$ and
$A(x\to \infty)
=- i \partial U(x) U^{-1}(x)$
where $U(x) \in SU(2)$.
The above definition includes configurations with odd gauge fields
$A_{i}(x)=-A_i(-x)$ such as the
S sphaleron\cite{KM}, the S$^*$ new sphaleron\cite{Klin}
and multisphaleron configurations\cite{Kunz}. It
further implies their classification in terms of the Chern-Simons number
functional
\be
N_{cs}(A) = \frac {1}{8 \pi^2}\int_{D^3} {\rm Tr}\,
(A d A - \frac{2 i}{3} A^3)\ .
\ee
Configurations with $U(x) = U(-x)$ (even $U$-parity) on the
two-sphere at infinity are found\cite{MA} to possess an integer
\mbox{$N_{cs}=n\in{\bf Z}$.} They include the S$^*$ new
sphaleron\cite{Klin}
with $N_{cs}=0$ and
multisphalerons with $N_{cs}=n \in {\bf Z}$ \cite{Kunz}. Configurations
with
$U(x) = - U(-x)$ (odd $U$-parity) at infinity have
$N_{cs}=n+1/2$. Examples are the S sphaleron with $N_{cs}=1/2$ and
multisphalerons with $N_{cs}=n+1/2$.

Such a classification stems from the existence of a ${\bf Z}_2$
topological structure
in the space of 3-dimensional static configurations in
$SU(2N)$, $SO(2N)$ and $E_7$
Yang-Mills theories\cite{Mol}. More precisely it is a consequence of the
existence of ${\bf Z}_2$ homotopy groups of maps $[ S^2/{\bf Z}_2,
G/{\bf Z}_2]={\bf Z}_2$ and
$[S^3/{\bf Z}_2, G/{\bf Z}_2] = {\bf Z} \times {\bf Z}_2$ .
Here $G= SU(2N)$, $SO(2N)$, or $E_{7}$.
Equivalently the Chern-Simons functional
restricted to odd-parity gauge fields is a topological charge. It takes
values in ${\bf Z}$ or ${\bf Z}+1/2$ depending on the even(odd) $U$-parity of
the underlying
gauge field configuration. As a consequence the
configuration space of the $SU(2) \times U(1)$
theory admits a ${\bf Z}_{2}$ structure. It consists
of two topologically disconnected classes of gauge-field configurations
which
are odd under the generalized parity transformation (Eq.(1)), itself
a composition of an ordinary parity and a gauge transformation.
An immediate implication of the above classification
is that the number of normalizable zero-energy solutions of the
3-dimensional Dirac
equation in the background of odd-parity gauge fields is a topological
invariant
modulo 2 in the Weinberg-Salam theory.
This is easy to see by considering a Dirac operator $\Dslash=
\sigma_i(\partial_i-iA_i)$ in an external odd-parity gauge field $A_{i}$.
It may be observed that its non-zero eigenvalues appear in pairs $(\lambda,
-\lambda)$. Indeed if $\psi(x)$ is a wave function that corresponds to an
eigenvalue $\lambda$ then $\psi(-x)$ is an eigenfunction corresponding to
an eigenvalue $-\lambda$. The argument generalizes trivially for the
case of an external $A_{i}$ field odd under a generalized parity as given
by equation (1).
Hence when the external field varies continuously
the number of zero modes of the Dirac operator is invariant modulo 2.
An odd $U$-parity gauge field background, such as the sphaleron,
provides
an odd number of these zero modes, thus allowing fermionic level
crossings which contribute
to anomalous $B$ violation in the Early Universe. In fact
there exists an infinite surface of gauge fields
with this property, in which the electroweak
sphaleron possesses the lowest energy $(E_{sp} \approx M_w/\alpha)$ and
electroweak string configurations\cite{MA} in the vicinity have energies
$E \geq E_{sp}$.

Even $U$-parity fields on the other hand give rise to an even number of
fermionic
zero modes. They are equivalent to the trivial vacuum and do not mediate
anomalous $B$ violation.
The S$^*$ new sphaleron with $N_{cs}=0$ is an example
as well as properly twisted electroweak string loops\cite{MA}.

In conclusion we have established that the Chern-Simons functional is
simultaneously a topological charge and an index for the 3-dimensional
Dirac operator.
This reflects the presence of a ${\bf Z}_2$
topological structure in the configuration
space of the bosonic sector of the hot electroweak theory. $U$-parity
provides a neat description of the skeleton of $B$-violating configurations
in the regime of temperatures where perturbation theory is valid.
We close with a few comments on the very concept of a skeleton we
alluded to.
While the ${\bf Z}_2$ topological structure provides us with an infinity of
gauge-Higgs field configurations with $B+L$ violating properties we cannot
claim to have exhausted the whole space of such configurations. In fact,
if we consider a path connecting two vacuum states with $N_{cs}=n ,n+1$
passing through
some odd-parity configuration with $N_{cs}=n+1/2$ it can be shown by
continuity of the Dirac spectrum that if a small even-parity gauge field
$A_E$ is added to every configuration along the path, satisfying $A_{E}(n)=
A_{E}(n+1)=0$, then an odd number of zero modes must occur for some
$ n < N_{cs} < n+1 $.

We will nevertheless argue that the space of generalized odd-parity
gauge-Higgs field
configurations with odd pure-gauge behavior at infinity constitutes the
backbone of the ``body'' of all configurations that mediate $B+L$ violation.
This is true in the broken phase of the theory $( M_{w}(T)\neq 0)$ where
the odd-parity saddle point alone\cite{Yaffe} governs the dynamics of
$B$-violating thermal transitions. As the temperature rises sphaleron
deformations with odd generalized parity are the first to become excited.
This can be seen through the consideration of a sphere $S_{E}$ in the
space E of sphaleron deformations which are orthogonal to the unstable
mode. On such a compact sphere the energy functional assumes maximal and
minimal values. As it is also invariant under the generalized parity
transformation of Eq.(1), odd-parity fields $A_{i}$ extremize
the energy on $S_{E}$.
By continuity such an extremum constitutes a minimum.

\section*{Acknowledgments}
The research of A.\,J. was partially supported by a NATO grant GRG 930395.
M.\,A.
acknowledges partial support from Danmarks Grundforskningsfond through
its contribution to the establishment of the Theoretical Astrophysics Center
as well as from the European Union under contract no.\ ERBCHRXCT 940621.

\end{document}